\documentstyle[11pt,amssymb]{article}
\newcommand{\be}{\begin{equation}}
\newcommand{\ee}{\end{equation}}
\newcommand{\ba}{\begin{array}}
\newcommand{\ea}{\end{array}}
\newcommand{\bea}{\begin{eqnarray}}
\newcommand{\eea}{\end{eqnarray}}

\renewcommand{\l}{\newline\null}

\newcommand{\rar}{\rightarrow}
\newcommand{\p}{\partial}
\newcommand{\ol}{\overline}
\newcommand{\ti}{\tilde}
\newcommand{\la}{\langle}
\newcommand{\ra}{\rangle}

\def\hbar{h\!\!\!/}

\begin{document}
\begin{titlepage}
June 1996\hfill PAR-LPTHE 96/21
%\begin{flushright} hep-ph/9606308 \end{flushright}
\vskip 4cm
\centerline{\bf LEPTONIC CUSTODIAL SYMMETRY,}
\centerline{\bf QUANTIZATION OF THE ELECTRIC CHARGE}
\centerline{\bf AND THE NEUTRINO IN THE STANDARD MODEL.}

\vskip .5cm
\centerline{B. Machet
     \footnote[1]{Member of `Centre National de la Recherche Scientifique'.}
     \footnote[2]{E-mail: machet@lpthe.jussieu.fr.}
           }
\vskip 5mm
\centerline{{\em Laboratoire de Physique Th\'eorique et Hautes Energies,}
     \footnote[3]{LPTHE tour 16\,/\,1$^{er}\!$ \'etage,
          Universit\'e P. et M. Curie, BP 126, 4 place Jussieu,
          F 75252 PARIS CEDEX 05 (France).}
}
\centerline{\em Universit\'es Pierre et Marie Curie (Paris 6) et Denis
Diderot (Paris 7);} \centerline{\em Unit\'e associ\'ee au CNRS URA 280.}
\vskip 1.5cm
{\bf Abstract:}  
I study, in the leptonic sector, the role, 
of the $SU(2)_V$ custodial symmetry $\ti{\cal G}$ which was shown 
in ref.~\cite{Machet} to control the quantization of the electric charge 
in the $J=0$ mesonic sector. The electroweak theory is considered, according
to ref.~\cite{BellonMachet}, as a purely vectorial model which interacts
with a ``hidden'' sector of composite scalars. $\ti{\cal G}$ can only be a
symmetry of the former if the neutrino is a Majorana particle; the latter
provides a dynamical modification of the leptonic weak couplings,
reconstructing those of the Standard Model with a massless Majorana
neutrino.
\smallskip

% \centerline{\rule {3cm} {0.2pt}}
\vfill
%\null\hfil\epsffile{/users/lpthe/machet/Papiers/logo/LogoCNRS.ps}
\end{titlepage}
\section{Introduction.}

The eigenvalues of the electric charge operator are naturally quantized if 
it is one of the generators of a simple group of symmetry of
the  theory under consideration; it can indeed always be then considered 
as the ``$z$'' component of an $SU(2)$ angular momentum subgroup.

In \cite{Machet}, I constructed a renormalizable, anomaly-free, gauge 
$SU(2)_L \times U(1)$ electroweak theory for $J=0$ mesons by extending the
scalar sector of the  Standard Model \cite{GlashowSalamWeinberg}, and showed
that it has, classically, independently of the hypercharge coupling 
constant $g'$, 
a global  ``custodial'' $SU(2)_V$ symmetry  which includes 
the $U(1)$ group of electromagnetism; it is the (unbroken) diagonal subgroup
of the chiral $SU(2)_L \times SU(2)_R$.

The underlying suggestion there is that the quantization of
the electric charge in this sector is the reflection that this
custodial $SU(2)_V$  is an exact symmetry of the model.
The fitting of each complex doublet of mesons into one $SU(2)_V$ real
triplet with electric charges $(-1, 0, +1)$  plus one real chargeless 
singlet eases the connection between this group and its electromagnetic
subgroup.

The case of leptons appears less intuitive since each family, in the
Glashow-Salam-Weinberg model \cite{GlashowSalamWeinberg} is 
cast into one doublet plus one singlet of the underlying $SU(2)_L$ gauge
group of symmetry, both including a charged particle;
there is clearly a mismatch here between their integer electric
charges and half-integer values of $SU(2)_L$ ``$z$'' quantum numbers.

We nevertheless adopt the unifying  point of view that the leptonic 
quantization of electric charge is the sign that a custodial $SU(2)$
symmetry also operates for asymptotic states.

This sector is studied here  in relation with  ref.~\cite{BellonMachet}. 
It is described by a purely vectorial theory which
rebuilds the Standard Model by interacting with a
``hidden'' sector.  We are concerned whether the fundamental vectorial
theory for the leptonic asymptotic states can satisfy charge quantization,
that is whether it can be globally invariant by the same custodial $SU(2)$ as
that of the mesonic sector.  We do not require this invariance for the whole
electroweak theory since the ``hidden'' sector does not correspond to
asymptotic states.

\section{Groups and representations.}

In \cite{Machet}, I introduced two commuting $SU(2)$ groups
\footnote{called there ``left'' and ``right'', but we denote them
differently here to avoid confusion.}
${\cal G}_1$ and ${\cal G}_2$, with generators $\vec{\Bbb T}_1$ and $\vec{\Bbb
T}_2$, which acted on special (quadruplet) representations of $J=0$ mesons
\be
\left( {\Bbb M}^0, \vec{\Bbb M} \right)
\ee
as follows (up to an irrelevant global sign):\l
- on representations ``even'' by the parity changing operator $\cal P$
(scalar $+$ pseudoscalar):

\vbox{
\bea
{\Bbb T}^i_1 . {\Bbb M}^j &=& 
                      {i\over 2}(\epsilon_{ijk}{\Bbb M}^k +
                                \delta_{ij}{\Bbb M}^0), \cr
{\Bbb T}^i_1 . {\Bbb M}^0 &=& -{i\over 2} {\Bbb M}^i,
\label{eq:G1}\eea
}

and

\vbox{
\bea
{\Bbb T}^i_2 . {\Bbb M}^j &=& 
                              {i\over 2}(\epsilon_{ijk}{\Bbb M}^k -
                                \delta_{ij}{\Bbb M}^0), \cr
{\Bbb T}^i_2 . {\Bbb M}^0 &=& {i\over 2} {\Bbb M}^i.
\label{eq:G2}\eea
}
- on representations ``odd'' (scalar $-$ pseudoscalar), the roles of the two
groups are swapped: ${\cal G}_1$ acts on ``odd'' representations like ${\cal
G}_2$ on ``even'' representations, and vice-versa.

The dot ``$.$'' between a generator and a field $\Bbb M$ stands for the 
action of the former on the latter.
The Latin indices $i,j,k$ run from $1$ to $3$, and we also define
${\Bbb M}^\pm = ({\Bbb M}^1 \pm i{\Bbb M}^2) /\sqrt{2}$,
${\Bbb T}^\pm = {\Bbb T}^1 \pm i {\Bbb T}^2$.

The quadruplets are reducible representations of each of these two groups,
and can be
decomposed into two spin $1/2$ doublets. With respect to the diagonal
$SU(2)_V$ subgroup $\ti{\cal G}$ of the chiral group 
${\cal G}_1 \times {\cal G}_2$, they
decompose differently, into one spin $1$ triplet, $\vec{\Bbb M}$, plus one
singlet ${\Bbb M}^0$. $\ti{\cal G}$ is the custodial symmetry which 
occurs in the mesonic sector.
The generator of the $U(1)$ group of electromagnetism is the ``$z$'' generator
of this angular momentum.

Let us now consider, for each generation of leptons, the left-handed 
\footnote{We shall use hereafter the subscript ``$L$'' for ``left-handed'' 
          fermions, and ``$R$'' for ``right-handed'' ones.}
quadruplet of neutrinos and charged leptons
(we do not question $e-\mu-\tau$ universality
and all leptons are {\em a priori} 4-components (chiral) leptons)

\be
{\cal Q}_L = \left({\Bbb L}^0, {\Bbb L}^3, {\Bbb L}^+, {\Bbb L}^- \right)
         =  \left(
         -i{\nu -\nu^c \over\sqrt{2}}, {\nu + \nu^c \over\sqrt{2}},
          \ell^+, \ell^-
                                  \right)_L.
\ee
$\ell^+$ and $\ell^-$, $\nu^c$ and $\nu$ are charge conjugate:
\be
\ell^+ = C\ol{\ell^-}^T, \quad \nu^c = C{\ol\nu}^T;
\ee
the superscript ``$T$'' means ``transposed'' and $C$ is the charge-conjugation 
operator: $C = i\gamma_2\gamma_0$ in the Dirac representation.
The convention that $\ell^+ = (\ell^1 + i \ell^2)/\sqrt{2}$ is the charge 
conjugate of $\ell^- = (\ell^1 - i \ell^2)/\sqrt{2}$ entails that 
$i$ gives $-i$ by charge conjugation
and that the charge conjugate of ${\cal Q}_L$ is its right-handed
counterpart ${\cal Q}_R = ({\Bbb R}^0, \vec{\Bbb R})$.

We define the actions of ${\cal G}_1$ and ${\cal G}_2$ on ${\cal Q_L}$
by eqs.~(\ref{eq:G1}) and (\ref{eq:G2}). 

${\cal Q}_L$ can be decomposed into:\l
- two doublets of ${\cal G}_1$:
\be
l_{1} = \left( \ba{c}  {1\over\sqrt{2}}({\Bbb L}^3 + i {\Bbb L}^0) \cr
                    {\Bbb L}^-
     \ea \right)
    = \left( \ba{c} \nu \cr
                    \ell^-
     \ea \right)_L,\qquad
l'_{1} = \left( \ba{c} {\Bbb L}^+ \cr
                   {1\over\sqrt{2}}({\Bbb L}^3 - i {\Bbb L}^0)
      \ea \right)
    = \left( \ba{c} \ell^+ \cr
                    \nu^c
       \ea \right)_L,
\label{eq:G1doublets}\ee
with the group action

\vbox{
\bea
& &{\Bbb T}_1^3.\ell_L^- = - {1\over 2} \ell_L^-,\quad
{\Bbb T}_1^3.\ell_L^+ = {1\over 2} \ell_L^+,\quad
{\Bbb T}_1^3.\nu_L = {1\over 2}\nu_L,\quad
{\Bbb T}_1^3.(\nu^c)_L = -{1\over 2}(\nu^c)_L,\cr
& &{\Bbb T}_1^+.\ell_L^- = \nu_L,\quad
{\Bbb T}_1^+.\ell_L^+ = 0,\quad
{\Bbb T}_1^+.\nu_L = 0,\quad
{\Bbb T}_1^+.(\nu^c)_L = -\ell_L^+,\cr
& &{\Bbb T}_1^-.\ell_L^- = 0,\quad
{\Bbb T}_1^-.\ell_L^+ =-(\nu^c)_L,\quad
{\Bbb T}_1^-.\nu_L = \ell_L^-,\quad
{\Bbb T}_1^-.(\nu^c)_L = 0;
\label{eq:G1action}\eea
}

${\cal G}_1$ acts on $l_{1}$ like the $SU(2)_L$ group of the Standard
Model.

- two doublets of ${\cal G}_2$:
\be
    l_{2} = \left( \ba{c} \nu^c \cr
                    \ell^-
     \ea \right)_L,\qquad
    l'_{2} =  \left( \ba{c} \ell^+ \cr
                    \nu
       \ea \right)_L,
\ee
with the group action

\vbox{
\bea
& &{\Bbb T}_2^3.\ell_L^- = -{1\over 2} \ell_L^-,\quad
{\Bbb T}_2^3.\ell_L^+ = {1\over 2} \ell_L^+,\quad
{\Bbb T}_2^3.\nu_L = -{1\over 2}\nu_L,\quad
{\Bbb T}_2^3.(\nu^c)_L = {1\over 2}(\nu^c)_L,\cr
& &{\Bbb T}_2^+.\ell_L^- = (\nu^c)_L,\quad
{\Bbb T}_2^+.\ell_L^+ = 0,\quad
{\Bbb T}_2^+.\nu_L = -\ell_L^+,\quad
{\Bbb T}_2^+.(\nu^c)_L = 0,\cr
& &{\Bbb T}_2^-.\ell_L^- = 0,\quad
{\Bbb T}_2^-.\ell_L^+ =-\nu_L,\quad
{\Bbb T}_2^-.\nu_L = 0,\quad
{\Bbb T}_2^-.(\nu^c)_L = \ell_L^-;
\label{eq:G2action}\eea
}

- one triplet $({\Bbb L}^3, {\Bbb L}^+, {\Bbb L}^-)$ plus one 
singlet ${\Bbb L}^0$ of the
diagonal $SU(2)_V$ with generators 
$\ti{\Bbb T}^i = {\Bbb T}^i_1 + {\Bbb T}^i_2$, with the group action

\vbox{
\bea
& &\ti{\Bbb T}^3.\ell_L^- = -\ell_L^-,\quad
\ti{\Bbb T}^3.\ell_L^+ = \ell_L^+,\quad
\ti{\Bbb T}^3.\nu_L = 0,\quad
\ti{\Bbb T}^3.(\nu^c)_L = 0,\cr
& &\ti{\Bbb T}^+.\ell_L^- = \nu_L + (\nu^c)_L,\quad
\ti{\Bbb T}^+.\ell_L^+ = 0,\quad
\ti{\Bbb T}^+.\nu_L = -\ell_L^+,\quad
\ti{\Bbb T}^+.(\nu^c)_L = -\ell_L^+,\cr
& &\ti{\Bbb T}^-.\ell_L^- = 0,\quad
\ti{\Bbb T}^-.\ell_L^+ = -(\nu_L + (\nu^c)_L),\quad
\ti{\Bbb T}^-.\nu_L = \ell_L^-,\quad
\ti{\Bbb T}^-.(\nu^c)_L = \ell_L^-.
\label{eq:GVaction}\eea
}

When operating in the 4-dimensional vector space spanned by the four entries
of ${\cal Q}_L$, its three generators  write as $4\times 4$ matrices
(see eq.~(28) of ref.~\cite{Machet}), according to (in the basis $({\Bbb
L}^0, {\Bbb L}^3, {\Bbb L}^+ {\Bbb L}^-)$):
\be
\tilde{\Bbb T}^+ = \left( \ba{cccc}
                   0  &  0        &  0        &  0  \cr
                   0  &  0        & -\sqrt{2} &  0  \cr
                   0  &  0        &  0        &  0  \cr
                   0  &  \sqrt{2} &  0        &  0
\ea\right),\quad
\tilde{\Bbb T}^- = \left( \ba{cccc}
                   0  &  0        &  0       &  0   \cr
                   0  &  0        &  0       &  \sqrt{2} \cr
                   0  & -\sqrt{2} &  0       &  0 \cr
                   0  &  0        &  0       &  0
\ea\right),\quad
\tilde{\Bbb T}^3 = \left( \ba{cccc}
                   0  &  0        &  0       &  0  \cr
                   0  &  0        &  0       &  0  \cr
                   0  &  0        &  1       &  0  \cr
                   0  &  0        &  0       & -1
\ea\right);
\label{eq:SU2V}\ee

The electric charge generator is identical with the third generator of
$SU(2)_V$:
\be
{\Bbb Q} = \ti{\Bbb T}^3.
\ee
By operating with charge conjugation on the transformed, by an
element of the group, of an entry of ${\cal Q}_L$, one deduces that its
charge conjugate ${\cal Q}_R$  also transforms  according to eqs.~(\ref{eq:G1},
\ref{eq:G2}). The decompositions above thus  apply to ${\cal Q}_R$ too.

To each quadruplet of leptons is associated a quadratic scalar
expression invariant by all $SU(2)$ groups considered above
\be
{\cal I} = \ol{\cal Q} {\cal Q} 
     =  \ol{{\cal Q}_R} {\cal Q}_L + \ol{{\cal Q}_L} {\cal Q}_R .
\label{eq:invar}\ee

\section{Chiral symmetry and  vector-like Lagrangians for leptons.}

We intend to make the link between a vectorial Lagrangian like that in
\cite{BellonMachet} and a chiral symmetry ${\cal G}_1 \times {\cal G}_2$
like that occurring naturally in \cite{Machet}. I show here that such a 
vectorial Lagrangian can be considered as a ${\cal G}_1 \times U(1)$ 
(or ${\cal G}_2 \times U(1)$) gauge theory for ${\cal Q}_L$ or ${\cal Q}_R$.
The chiral symmetry then naturally springs out at the limit when the two
electroweak couplings vanish.

Let us construct a ${\cal G}_1 \times U(1)$ gauge theory for
${\cal Q}_L$. The three ${\cal G}_1$ gauge fields 
form the triplet $\vec W_\mu$, and the $U(1)$ gauge field is $B_\mu$.

The generator $\Bbb Y$ of the $U(1)$ group is taken to
satisfy the Gell-Mann-Nishijima relation
\be
{\Bbb Y} = {\Bbb Q} - {\Bbb T}^3_1.
\label{eq:GMN}\ee
The $U(1)$ is a gauging of the leptonic number; indeed, the leptonic 
numbers of the entries of $\cal Q$ are $(-2) \times$ their $U(1)$ quantum numbers.

The notations are, as usual
\be
Z_\mu = c_w W_\mu^3 - s_w B_\mu,\  A_\mu = c_w B_\mu + s_w W_\mu^3,
\ee
where $c_w$ and $s_w$ are the cosine and sine of the Weinberg angle; $A_\mu$
is the photon; $W_\mu^\pm = (W_\mu^1 \pm W_\mu^2)/\sqrt{2}$;
$g = e/s_w$ and $g' = e/c_w$ are respectively the ${\cal G}_1$ and
$U(1)$  coupling constants.

The Lagrangian  is the sum of two Lagrangians 
${\cal L}_1$ and ${\cal L}'_1$ corresponding
respectively to the two doublets $l_1$ and $l'_1$. We use the abbreviated
notation $\gamma_{\mu L} = \gamma_\mu (1-\gamma_5)/2, \gamma_{\mu R} =
\gamma_\mu(1+\gamma_5)/2$.

\vbox{
\bea
{\cal L}_1 = & &i \ol{\ell^-} \gamma^\mu_L \p_\mu \ell^-
                  + i \bar \nu \gamma^\mu_L \p_\mu \nu\cr
             &+& {e\over \sqrt{2}s_w}\left(
           \ol{\ell^-} \gamma^\mu_L  W_\mu^- \nu
           + \ol\nu \gamma^\mu_L  W_\mu^+ \ell^-
                             \right)\cr
             &-& {e\over 2s_w}\left(
                 \ol{\ell^-} \gamma^\mu_L W_\mu^3\ell^-
             - \ol\nu \gamma^\mu_L W_\mu^3\nu \right)\cr
             &-& {e\over 2c_w} \left( \ol{\ell^-} \gamma^\mu_L  B_\mu \ell^-
                                 + \ol\nu \gamma^\mu_L B_\mu \nu \right);
\label{eq:L1}
\eea
}

\vbox{
\bea
{\cal L}'_1 = & &i \ol{\ell^+} \gamma^\mu_L \p_\mu \ell^+
                  + i \ol{\nu^c} \gamma^\mu_L \p_\mu \nu^c\cr
             &-& {e\over \sqrt{2}s_w}\left(
            \ol{\ell^+} \gamma^\mu_L  W_\mu^+ \nu^c
       -    \ol{\nu^c} \gamma^\mu_L  W_\mu^- \ell^+
                             \right)\cr
             &+& {e\over 2s_w}\left(
                 \ol{\ell^+} \gamma^\mu_L W_\mu^3\ell^+
             - \ol{\nu^c} \gamma^\mu_L W_\mu^3\nu^c \right)\cr
             &+& {e\over 2c_w} \left( \ol{\ell^+} \gamma^\mu_L B_\mu \ell^+
                          + \ol{\nu^c} \gamma^\mu_L B_\mu \nu^c \right).
\label{eq:L'1}
\eea
}
Using the properties of charge conjugation in ${\cal L}'_1$, ${\cal L}_1 + 
{\cal L}'_1$ can be cast into the purely vectorial Lagrangian of 
\cite{BellonMachet}; it is a $SU(2) \times U(1)$ gauge Lagrangian for a
$(\nu, \ell^-)$ doublet:

\vbox{
\bea
{\cal L} = {\cal L}_1 + {\cal L}'_1 = 
              & &i \ol{\ell^-} \gamma^\mu \p_\mu \ell^- 
                  + i \bar \nu \gamma^\mu \p_\mu \nu\cr
           &+& {e\over \sqrt{2}s_w}\left(
           \ol{\ell^-} \gamma^\mu  W_\mu^- \nu
      +    \ol\nu \gamma^\mu  W_\mu^+ \ell^-
                             \right)\cr
             &-& {e\over 2s_w}\left(
                 \ol{\ell^-} \gamma^\mu W_\mu^3\ell^-
             - \ol\nu \gamma^\mu W_\mu^3\nu \right)\cr
             &-& {e\over 2c_w} \left( \ol{\ell^-} \gamma^\mu  B_\mu \ell^-
                               + \ol\nu \gamma^\mu  B_\mu \nu\right).
\label{eq:L}
\eea
}

The same result can be obtained by considering ${\cal Q}_R$ instead of
${\cal Q}_L$.

To $\cal L$ we can add the mass term
\be
{\cal L}_m 
=   -{m\over 2} \left( \ol{{\cal Q}_R} {\cal Q}_L + \ol{{\cal Q}_L} {\cal
Q}_R \right).
\label{eq:Lm}\ee
The quadratic expression (\ref{eq:invar}) being invariant by both ${\cal
G}_1$ and ${\cal G}_2$, ${\cal L}_m$ is invariant by the chiral group ${\cal
G}_1 \times {\cal G}_2$, and
this invariance is independent of the mass $m$, which can vary with the
leptonic generation.

${\cal L}_m$ corresponds to a Dirac mass term. It is an important actor in the
``see-saw'' mechanism evoked in the last section. A Majorana mass term for the
neutrino would correspond to the combination (forgetting the charged
leptons) $(\ol{{\Bbb R}^3} {\Bbb L}^3 + \ol{{\Bbb L}^3} {\Bbb R}^3) 
- (\ol{{\Bbb R}^0} {\Bbb L}^0 + \ol{{\Bbb L}^0} {\Bbb R}^0) + \cdots$,
(the ``$-$'' sign makes the difference), which is not invariant by ${\cal G}_1$.

Would we make a similar construction with the gauge group ${\cal G}_2 \times
U(1)$, we would obtain a Lagrangian similar to (\ref{eq:L}) but with $\nu$
and $\nu^c$ swapped. As this leaves the kinetic terms unaltered, ${\cal L} + 
{\cal L}_m$ has a global ${\cal G}_1 \times {\cal G}_2$ chiral symmetry at 
the limit $g,g' \rightarrow 0$, hence also a global symmetry by the diagonal
$\ti{\cal G}$.

\section{The custodial \boldmath{$SU(2)_V$} symmetry.}

Non-vanishing values of $g$ and $g'$ break the ${\cal G}_1 \times {\cal
G}_2$ chiral symmetry. We ask at which condition the diagonal group $\ti{\cal
G}$ can stay an unbroken global  symmetry of ${\cal L} + {\cal L}_m$,
providing an understanding of the quantization of the electric charge for the
electroweak vector-like model under scrutiny, with the same origin as in
the mesonic sector \cite{Machet}.

We can rewrite ${\cal L}_1 + {\cal L}'_1$ in the form
\bea
& &2({\cal L}_1 + {\cal L}'_1) =  \cr
 & &i(\ol{{\Bbb L}^+} \gamma_\mu \p^\mu {\Bbb L}^+
                +  \ol{{\Bbb L}^-} \gamma_\mu \p^\mu {\Bbb L}^-
                +  \ol{{\Bbb L}^3} \gamma_\mu \p^\mu {\Bbb L}^3
                +  \ol{{\Bbb L}^0} \gamma_\mu \p^\mu {\Bbb L}^0)\cr
&+&{g} \left(
\ol{{\Bbb L}^-} \gamma_{\mu L} ({1\over \sqrt{2}}
                      (W_\mu^- \tilde{\Bbb T}^+ +W_\mu^+ \tilde{\Bbb T}^-)
                            +{Z_\mu \over c_w} \tilde{\Bbb T}^3).{\Bbb L}^- 
\right.\cr
& & \left. \hphantom{aaaaaaaaaaaaaaaaaaaaaaaaaaaaa}
+\ol{{\Bbb L}^+} \gamma_{\mu L} ({1\over \sqrt{2}}
                      (W_\mu^- \tilde{\Bbb T}^+ +W_\mu^+ \tilde{\Bbb T}^-)
                            +{Z_\mu \over c_w} \tilde{\Bbb T}^3).{\Bbb L}^+
\right.\cr
& & \left. \hphantom{aaaaaaaaaaaaaaaaaaaaaaaaaaaaa}
 +\ol{{\Bbb L}^3} \gamma_{\mu L} ({1\over \sqrt{2}}
   (W_\mu^- \tilde{\Bbb T}^+ + W_\mu^+ \tilde{\Bbb T}^-) 
     +{Z_\mu \over c_w} \tilde{\Bbb T}^3). {\Bbb L}^3
               \right)\cr
&+& i\;{g} \left(
  - \ol{{\Bbb L}^0} (\gamma_{\mu L} {Z_\mu \over c_w} {\Bbb L}^3
                    + \gamma_{\mu L} W_\mu^- {\Bbb L}^+
                    + \gamma_{\mu L} W_\mu^+ {\Bbb L}^-) \right. \cr
& & \left. \hphantom{aaaaaaaaaaaaaaaaaaaaaaaaaaaaa}  
           + (\ol{{\Bbb L}^3} \gamma_{\mu L} {Z_\mu \over c_w} 
        +\ol{{\Bbb L}^-} \gamma_{\mu L} W_\mu^-
        + \ol{{\Bbb L}^+} \gamma_{\mu L} W_\mu^+){\Bbb L}^0 
               \right)\cr
&+& {g'} (\ol{{\Bbb L}^+} \gamma_{\mu L}  B^\mu {\Bbb L}^+ 
           -\ol{{\Bbb L}^-} \gamma_{\mu L}  B^\mu {\Bbb L}^-),
\label{eq:Lag}\eea
where we have used the fact that $\tilde{\Bbb T}^+$ does not act on ${\Bbb
L}^+$, nor $\tilde{\Bbb T}^-$ on ${\Bbb L}^-$, nor $\tilde{\Bbb T}^3$ on 
${\Bbb L}^3$.

The kinetic terms and the second line of (\ref{eq:Lag})
are  globally $SU(2)_V$ invariant when the triplet of gauge bosons
$W_\mu^\pm$ and $Z_\mu/c_w$ transform, according to \cite{Machet},
like a vector in the adjoint representation of this group.

The next line of (\ref{eq:Lag}) is also globally $SU(2)_V$ invariant,
since ${\Bbb M}^0$ and $\ol{{\Bbb M}^0}$ are singlets and are each
multiplied by another singlet made by the scalar product of two triplets.

Now, $B_\mu$ being considered \cite{Machet} as a singlet of $SU(2)_V$,
the last line of (\ref{eq:Lag}) only becomes $SU(2)_V$ 
invariant if, as can be seen by performing an
explicit transformation and using (\ref{eq:GVaction},\ref{eq:SU2V})
\be
\nu  +\nu^c = 0,
\label{eq:Majorana}
\ee
{\em i.e.} the neutrino has to be a Majorana particle, 
with only one helicity (or
chirality), which can be written \cite{Ramond}, in the 4-component notation,
either $\gamma_5\chi$ or  $\gamma_5\omega$ with
\be
\chi = \left( \ba{c} \psi_L \cr
                      -\sigma^2 \psi_L^\ast \ea \right),\quad
\omega = \left( \ba{c} \psi_R \cr
                      -\sigma^2 \psi_R^\ast \ea \right).
\label{eq:chiomega}\ee
$\psi_L$ (resp. $\psi_R$)  is a two-component Weyl spinor transforming like 
a $(1/2,0)$ (resp. $(0,1/2)$) representation of the Lorentz group;
$\sigma^2$ is the second Pauli matrix and the superscript ``$\ast$'' means
``complex conjugation''; $\sigma^2 \psi_L^\ast$ (resp. $\sigma^2
\psi_R^\ast$) transforms like a $(0,1/2)$ (resp. $(1/2,0)$) representation.

We thus conclude that:\l
{\em The leptonic Lagrangian ${\cal L}$ can have a global custodial
$SU(2)_V$ symmetry only if  the neutrino is a Majorana particle.}

Clearly, this condition is not compatible with the decomposition
(\ref{eq:G1doublets}) and the corresponding laws of transformation 
(\ref{eq:G1}). In
particular, it requires that the $U(1)$ leptonic number be not conserved.
We shall see in the next section how the necessary modifications can occur
dynamically with the introduction of a ``hidden'' sector, along the lines of
\cite{BellonMachet}.

Like in \cite{Machet}, we now ask whether  the custodial
symmetry can become a {\em local} symmetry, with the triplet $W_\mu^\pm,
Z_\mu/c_w$ transforming like the corresponding gauge potentials.
The first two lines of (\ref{eq:Lag}) have been intentionally written as
those of an  $SU(2)_V$ gauge theory and are thus naturally locally invariant.

By such a local transformation with parameters $\vec\zeta(x)$, 
the third line varies by
\be
\delta = \vec J^\mu {\cal D}_\mu \vec\zeta,
\ee
where ${\cal D}_\mu$ is the covariant derivative with respect to $SU(2)_V$
\be
 {\cal D}_\mu {\Bbb L} = \p_\mu {\Bbb L} -ig \left(
       {1\over\sqrt{2}}(W_\mu^+ \tilde{\Bbb T}^- + W_\mu^- \tilde{\Bbb T}^+) 
                + {Z_\mu\over c_w}\tilde{\Bbb T}^3 \right). {\Bbb L}\,,
\ee
and $\vec J_\mu$ is a triplet of currents
\be
J_\mu^i = \ol{{\Bbb L}^i} \gamma^\mu_L {\Bbb L}^0 
      - \ol{{\Bbb L}^0}\gamma^\mu_L {\Bbb L}^i,\quad i = 1,2,3.
\ee
These currents are covariantly {\em with respect to ${\cal G}_1 \times
U(1)$} conserved.
As they form a triplet of $SU(2)_V$ and because the corresponding $J_\mu^0$
identically vanishes by definition, one has
\be
D^\mu J_\mu^i = {\cal D}^\mu J_\mu^i -ig' B_\mu \tilde{\Bbb Q}.J_\mu^i,
\ee
such that the invariance of the third line of (\ref{eq:Lag}),
requiring the vanishing of $\delta$ up to a total divergence, can only occur
when
\be
 g' B_\mu \tilde{\Bbb Q}.J_\mu^i =0, \quad i=1,2,3,
\ee
that is when $g' \rightarrow 0$.

At this limit, the last line of (\ref{eq:Lag}), which vanishes, 
is also trivially invariant.

We thus find the same result as in \cite{Machet} for $J=0$ mesons:\l
{\em The custodial $SU(2)_L$ symmetry becomes local in the leptonic sector
at the limit $g' \rightarrow 0$.}

\section{Left-right spontaneous symmetry breaking; back to the Standard
Model for leptons.}

I now summarize and comment the procedure of \cite{BellonMachet}, by which
the vector-like model (\ref{eq:L}) can be reconciled with the 
Glashow-Salam-Weinberg Lagrangian for leptons.
 
As Majorana neutrinos are involved, it is natural to introduce the notation
\bea
\chi &=& \nu_L + (\nu_L)^c,\cr
\omega &=& \nu_R + (\nu_R)^c;
\eea
and to rewrite the Lagrangian (\ref{eq:L}) $+$ (\ref{eq:Lm}) as:
\bea
{\cal L} + {\cal L}_m &=&
 i \ol{\ell^-} \gamma^\mu \p_\mu \ell^-
+ {i\over 2} \bar\chi \gamma^\mu \p_\mu \chi
+ {i\over 2} \bar\omega \gamma^\mu \p_\mu \omega\cr
&+& {e\over\sqrt{2} s_w}(\ol{\ell^-} \gamma^\mu_L W_\mu^- \chi 
                   +\bar\chi \gamma^\mu_L W_\mu^+ \ell^-)\cr
&+& {e\over\sqrt{2} s_w}(\ol{\ell^-} \gamma^\mu_R W_\mu^- \omega
                   +\bar\omega \gamma^\mu_R W_\mu^+ \ell^-)\cr
&+& {e\over 2 s_w}\left(
\bar\chi \gamma^\mu_L W_\mu^3\chi
+  \bar\omega  \gamma^\mu_R  W_\mu^3\omega
  -\ol{\ell^-} \gamma^\mu W_\mu^3 \ell^- \right)\cr
&-& {e\over 2c_w} \left(\ol{\ell^-} \gamma^\mu B_\mu \ell^-
    + \bar\chi \gamma^\mu_L B_\mu \chi
    + \bar\omega \gamma^\mu_R B_\mu \omega \right)\cr
&-& {m\over 2}(\bar\chi \omega + \bar\omega \chi + 2 \ol{\ell^-} \ell^-).
\label{eq:Lmajo}\eea
{\em Remark:} would we have built the model with the group ${\cal G}_2
\times U(1)$, we would have obtained, instead of eq.~(\ref{eq:Lmajo})
$\hat{\cal L}$, deduced from $\cal L$ by the exchange of $\chi$ and
$\omega$, or, equivalently, by that of the ``left'' and ``right'' projectors.

We introduce a scalar composite triplet $\Delta$ with leptonic number $2$:
\be
\Delta = \left( \ba {l} \Delta^{0} \cr
                        \Delta^{-} \cr
                        \Delta^{--} \cr
         \ea \right)
       ={\rho\over\nu^3}
          \left( \ba {c} \ol{\omega_L} \omega_R \cr
                          {1\over \sqrt{2}}(\ol{\ell^+_L} \omega_R+
                                    \ol{\omega_L} \ell^-_R)\cr
                          \ol{\ell^+_L} \ell^-_R
          \ea \right)
       = {\rho\over\nu^3}
          \left( \ba {c} \ol{\nu^c}\ {1+\gamma_5\over 2}\ \nu \cr
                    {1\over \sqrt{2}}(\ol{\ell^+}\ {1+\gamma_5\over 2}\ \nu +
                               \ol{\nu^c}\ {1+\gamma_5\over 2}\ \ell^-) \cr
                          \ol{\ell^+}\ {1+\gamma_5\over 2}\ \ell^-
          \ea \right).
\ee
It is a triplet of ${\cal G}_1$ but not a representation of ${\cal G}_2$,
nor of $\ti{\cal G}$.

Its hermitian conjugate is:
\be
\ol\Delta = \left( \ba {l} {\ol{\Delta^0}} \cr
                            \Delta^{+} \cr
                            \Delta^{++} \cr
         \ea \right)
       ={\rho\over\nu^3}
          \left( \ba {c} \ol{\omega_R} \omega_L \cr
                          {1\over \sqrt{2}}(\ol{\ell^-_R} \omega_L +
                                   \ol{\omega_R} \ell^+_L) \cr
                          \ol{\ell^-_R} \ell^+_L
          \ea \right)
       = {\rho\over\nu^3}
          \left( \ba {c} \ol{\nu}\ {1-\gamma_5\over 2}\ \nu^c \cr
                 {1\over \sqrt{2}}(\ol{\ell^-}\ {1-\gamma_5\over 2}\ \nu^c +
                                \ol{\nu}\ {1-\gamma_5\over 2}\ \ell^+) \cr
                          \ol{\ell^-}\ {1-\gamma_5\over 2}\ \ell^+
          \ea \right).
\ee
As soon as the mass of $\omega$ is non vanishing, electroweak vacuum
fluctuations like described in fig.~1 of \cite{BellonMachet} can trigger
\be
\la\Delta^0\ra = \la\ol{\Delta^0}\ra = \rho.
\ee
The choice of an $\la\bar\omega \omega\ra$ condensate, breaking the symmetry
between $\chi$ and $\omega$, spontaneously breaks 
the ``left-right'' symmetry, or, equivalently, parity.

It could be thought arbitrary since the same type of vacuum
fluctuations can also {\em a priori} trigger $\la \bar\chi \chi \ra \not=
0$. However, the diagrams under consideration vanish with the mass of the
internal fermion. As, by the see-saw mechanism evoked below, an 
$\la \bar\omega \omega \ra$ condensate pushes the $\chi$ mass to $0$ at
the same time that is pushes the $\omega$ mass to $\infty$,
the $\la \bar\chi \chi \ra$ condensate is then automatically suppressed, and
vice-versa. This qualitative explanation forbids the coexistence of both
condensates.
 
The proposed mechanism can also be interpreted along the following lines: 
by expanding
$\gamma_\mu$ into $(\gamma_{\mu L} + \gamma_{\mu R})$, the vectorial
Lagrangian (\ref{eq:L}) can be considered to be that of an  
$SU(2)_L \times SU(2)_R \times U(1)$ gauge model for the doublet 
$(\nu, \ell^-)$; both $SU(2)$'s 
act the same way, with the ``left''and ``right'' gauge fields identified.
The ``Higgs'' multiplet $\Delta$ being a  triplet of $SU(2)_R$ 
and of $SU(2)_L$ with a non-vanishing leptonic number,  the
condensation of its neutral component spontaneously breaks both $SU(2)$'s,
and  $U(1)$.
 
For quantizing by the Feynman path integral, a Lagrangian of constraint is
introduced to take into account the non-independence of the leptonic and
$\Delta$ degrees of freedom ($\Lambda$ is an arbitrary mass scale):
\be\ba{lcl}
{\cal L}_c &=& \lim_{\beta\rar 0}
            -{\Lambda^2\over 2\beta}\left[
\left( \Delta^0 -{\rho\over\nu^3}\, \ol{\omega_L} \omega_R \right)
\left(\ol{\Delta^0}-{\rho\over\nu^3}\, \ol{\omega_R} \omega_L\right)
\right.\cr
& &+
\left.\left(\Delta^- -{1\over\sqrt{2}}{\rho\over\nu^3}(\ol{\omega_L} \ell^-_R +
\ol{\ell^+_L} \omega_R) \right)
\left(\Delta^+ -{1\over\sqrt{2}}{\rho\over\nu^3}(\ol{\ell^-_R} \omega_L +
\ol{\omega_R} \ell^+_L) \right)\right.\cr
& &+
\left.\left (\Delta^{--} -{\rho\over\nu^3}\, \ol{\ell^+_L} \ell^-_R\right)
\left(\Delta^{++} -{\rho\over\nu^3}\, \ol{\ell^-_R} \ell^+_L\right)
\right].
\label{eq:Lc}\ea\ee
It has two types of effects, obtained be resumming diagrams at
leading order in an expansion in inverse powers of the number of generations
and by building a reshuffled perturbation series:
\l
- it gives $\omega$, through an exact ``see-saw'' mechanism, an infinite
(Majorana) mass, leaving $\chi$ as the only asymptotic neutrino state,
identified with the observed one; $\chi$ is massless; the left
coupling of the charged $W$'s to $\chi$ and $\ell^-$ is the only one left
over;\l
- it modifies, through internal $\omega$ loops, the effective couplings of
the neutral $W_\mu^3$ and $B_\mu$ gauge bosons, rebuilding the ones of the 
Standard Model; internal loops do not alter the above mentioned charged $W^\pm$
couplings.

Finally, the right-handed $\ell^-_R$ effectively behaves like a $SU(2)_L$
singlet with twice the hypercharge of the left-handed one, and the customary
structure of weak currents is recovered, in both the charged and neutral
sectors. The $\Delta^0$ condensate spontaneously breaks the $SU(2)_R \times
U(1)$ symmetry down to the $U(1)_Y$ of weak hypercharge.

None of the components of $\Delta$ appears as an asymptotic state
(the constraints give them an infinite mass)
and we do not require electric charge quantization for them.
It is the same kind of implicit assumption that we made in the
mesonic sector where explaining charge quantization for quarks and their
underlying gauge theory (Quantum Chromodynamics) was not sought for.
So, we allow a Lagrangian which is not globally $\ti{\cal G}$ invariant in 
the hidden sector; this non-invariance is responsible for that of the 
Lagrangian of the Standard Model.

Only ${\cal L}_c$, which couples the latter to asymptotic states,
has a global $\ti{\cal G}$ invariance; it occurs as soon as the laws of
transformations of the components of $\Delta$  and those of the corresponding 
composite fields are the same.

The kinetic terms for the scalars are constructed on the same group of
symmetry ${\cal G}_1 \times U(1)$ as $\cal L$ itself.
$SU(2)_L$ is also broken by the condensation of $\Delta^0$, which
thus weakly contributes to the masses of the gauge fields. However, by
the decoupling of the weak hidden sector, the Goldstones of this broken 
symmetry align with the customary hadronic ones. 
This decoupling also motivates non introducing
other triplets of composite states, with $\omega$ replaced by $\chi$, since
they would not modify the result: as soon as only one type of 
condensate can occur, the non-condensing additional scalars would simply
fade away without any visible effect.

\section{Conclusion.}

The Glashow-Salam-Weinberg Lagrangian has no conspicuous symmetry which could 
explain charge quantization for leptons. It is a motivation to consider 
it as an effective theory, with an underlying more fundamental level at 
which should be sought an explanation for basic observed features of asymptotic 
states.

It may look going astray to the reader that, in the mesonic sector
\cite{Machet}, we left aside the gauge theory of quarks and gluons to directly 
investigate an electroweak theory of composite asymptotic states, and here, 
that we added a questionable hidden sector of composite states to a superbly 
working Standard Model; this maybe paying a heavy price for a modification 
which brings new unanswered questions.

On the other side, it is hardly possible to solve all problems at the same 
time, and one must temporarily accept that filling gaps digs holes elsewhere. 
In front of fundamental questions unsolved, like precisely the origin of 
the quantization of the electric charge, I disturbed for a while
the harmony of the construction to tackle the problem from another
point of view. Developments should follow in subsequent works.

\bigskip
\begin{em}
\underline {Acknowledgments}: the author thanks P. Fayet for comments and
suggestions.
\end{em}
\newpage\null
\begin{em}

\end{em}


\begin{thebibliography}{50}
%
\bibitem{Machet}
       B. MACHET: ``Chiral Scalar Fields, Custodial Symmetry in
                   Electroweak $SU(2)_L \times U(1)$, and the Quantization 
                   of the Electric Charge'',
                            preprint PAR-LPTHE 96/20, hep-ph/9606239.

\bibitem{BellonMachet}
        M. BELLON and B. MACHET: ``The Standard Model of Leptons as a Purely
                          Vectorial Theory'',  Phys. Lett. B 313 (1993) 341.

\bibitem{GlashowSalamWeinberg}
        S. L. GLASHOW: Nucl. Phys. 22 (1961) 579;\l
        A. SALAM: in ``Elementary Particle Theory: Relativistic Groups and
             Analyticity'' (Nobel symposium No 8), edited by N. Svartholm
             (Almquist and Wiksell, Stockholm 1968);\l
        S. WEINBERG: Phys. Rev. Lett. 19 (1967) 1224.

\bibitem{Sikivie}
        P. SIKIVIE, L. SUSSKIND, M. VOLOSHIN and V. ZAKHAROV: Nucl. Phys. B
                                 173 (1980) 189.
\bibitem{Ramond}
        P. RAMOND: ``Field Theory; a Modern Primer'', Frontiers in Physics,
                 Lecture Notes Series 51 (Benjamin/Cummings 1981).


%
\end{thebibliography}
\end{document}